\documentclass[sigconf]{acmart}

\usepackage{enumitem}

\newcommand{\AgenticAKM}{AgenticAKM}

\AtBeginDocument{%
  }

\copyrightyear{2026}
\acmYear{2026}
\setcopyright{cc}
\setcctype{by}
\acmConference[AGENT '26]{International Workshop on Agentic Engineering }{April 12--18, 2026}{Rio de Janeiro, Brazil}
\acmBooktitle{International Workshop on Agentic Engineering (AGENT '26), April 12--18, 2026, Rio de Janeiro, Brazil}
\acmPrice{}
\acmDOI{10.1145/3786167.3788416}
\acmISBN{979-8-4007-2399-5/2026/04}

\begin{document}

\title{AgenticAKM : Enroute to Agentic Architecture Knowledge Management}
\settopmatter{printacmref=false}

\author{Rudra Dhar}
\affiliation{%
  \institution{IIIT Hyderabad}
  \city{Hyderabad}
  \state{Telangana}
  \country{India}}
\email{rudra.dhar@research.iiit.ac.in}

\author{Karthik Vaidhyanathan}
\affiliation{%
  \institution{IIIT Hyderabad}
  \city{Hyderabad}
  \country{India}}
\email{karthik.vaidhyanathan@iiit.ac.in}

\author{Vasudeva Varma}
\affiliation{%
  \institution{IIIT Hyderabad}
  \city{Hyderabad}
  \country{India}}
\email{vv@iiit.ac.in}

\renewcommand{\shortauthors}{Dhar et al.}


\begin{abstract}
Architecture Knowledge Management (AKM) is crucial for maintaining current and comprehensive software Architecture Knowledge (AK) in a software project. However AKM is often a laborious process and is not adopted by developers and architects. While LLMs present an opportunity for automation, a naive, single-prompt approach is often ineffective, constrained by context limits and an inability to grasp the distributed nature of architectural knowledge. To address these limitations, we propose an Agentic approach for AKM, \AgenticAKM, where the complex problem of architecture recovery and documentation is decomposed into manageable sub-tasks. Specialized agents for architecture Extraction, Retrieval, Generation, and Validation collaborate in a structured workflow to generate AK. To validate we made an initial instantiation of our approach to generate Architecture Decision Records (ADRs) from code repositories. We validated our approach through a user study with 29 repositories. The results demonstrate that our agentic approach generates better ADRs, and is a promising and practical approach for automating AKM.
\end{abstract}





\maketitle

\section{Introduction}\label{sec:introduction}

Software architecture provides the foundational blueprint of a system, but keeping its documentation accurate and complete remains a challenge. This documentation, which captures key Architectural Knowledge (AK) such as structure, components, and design principles, is vital for long-term maintenance and evolution. Effective Architecture Knowledge Management (AKM) ensures this knowledge is systematically created and preserved, thus reducing knowledge vaporization.

Despite its importance, the manual creation and maintenance of AK remain a significant bottleneck. This is especially true for crucial artifacts like Architecture Decision Records (ADRs), which capture Architectural Design Decisions. Documentation is often neglected, leading to records that are incomplete, outdated, or disconnected from implementation. This knowledge gap makes it difficult for teams to understand the system's design, onboard new developers, and make informed decisions during its evolution.

The recent advancements in Large Language Models (LLM) present a opportunity to automate AKM by analyzing code repositories and generate relevent documentation. However, a naive single prompt approach, like feeding an entire repository to an LLM is often ineffective and is constrained by LLM's context window, and results in outputs that are inaccurate, or lacking essential context.

To overcome these challenges, we propose an agentic approach, a paradigm gaining significant traction for complex tasks \cite{xi2025rise}, including within software engineering \cite{he2025llm, wang2025agents}.
While Sapkota et al. \cite{sapkota2025ai} attempts to define multi agent systems, frameworks like AutoGen \cite{wu2023autogenenablingnextgenllm} demonstrates the power of multi-agent collaboration. We adapt this concept to AKM, introducing an agentic approach, \AgenticAKM, where the complex problem of architecture recovery and documentation is decomposed. Our approach is built upon four distinct types of agents, each responsible for a key stage of the process: Architecture Extraction, Retrieval, Generation, and Validation. These agents with specific roles and tools, are coordinated by a central orchestrator.

In this paper, we present the design and an instantiation of \AgenticAKM. In the instantiation, \AgenticAKM~analyzes a code repository and produces a set of ADRs. We validate it through a user study comparing our approach against a baseline single LLM call. The results demonstrate that \AgenticAKM~produces significantly better ADRs establishing it as a promising and practical approach for automating AKM.

The rest of the paper is structured as follows. Section \ref{sec:motivation} gives an motivation and overview of the Agentic approach, whereas section \ref{sec:agents} explains the various agents. Section \ref{sec:experiments} details our experimentation. Finally section \ref{sec:RelatedWorks} discusses some related works, and section \ref{sec:conclusion} concludes the work.

\section{Motivation and Overview}\label{sec:motivation}

\begin{figure*}[ht]
    \centering
    \includegraphics[width=0.9\textwidth]
    {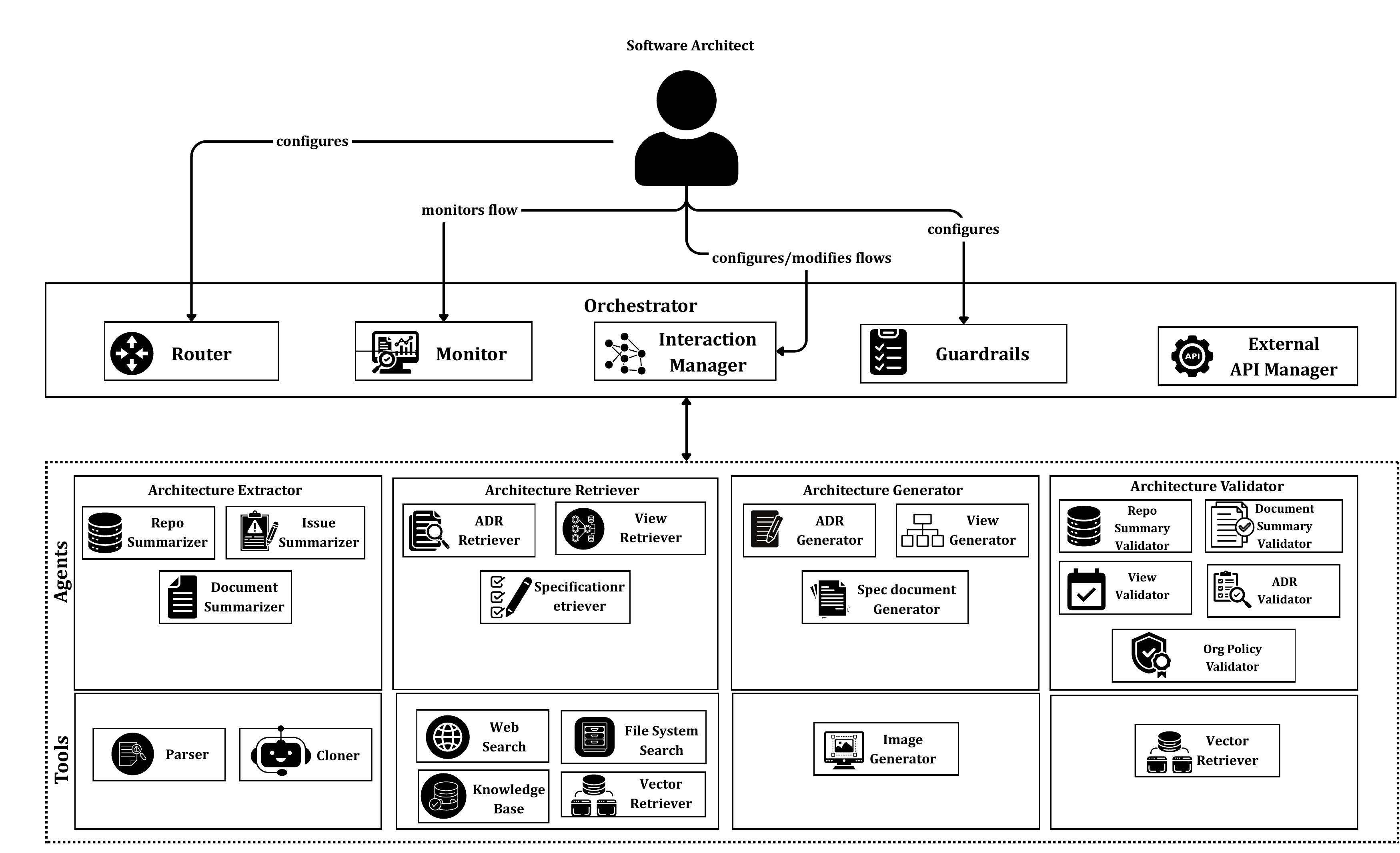}
    \caption{Agentic AKM}
    \label{fig:AgenticApproach}
\end{figure*}

Recent advances in LLMs offer opportunities to automate AKM by reasoning over code, configurations, and documentation to generate consistent, context-aware artifacts and reduce architects’ workload. However, AK is highly distributed and spans multiple abstraction levels beyond a single prompt’s capacity, while limited context windows hinder full codebase analysis.

LLM generated outputs often appear plausible but lack depth, omitting design rationales and historical context. Their opaque reasoning also hinders verification. For example, generating ADRs from a repository requires synthesizing information from code, commits, and issue trackers—something a single LLM call typically fails to achieve, resulting in incomplete or misleading documentation.

To overcome these challenges, we propose \AgenticAKM. Instead of relying on a single model, we employ a multi-agent approach where the complex problem is decomposed into distinct, manageable sub-tasks. Each task is handled by a specialized AI agent with a specific role and toolset, all coordinated by a central orchestrator as shown in Figure \ref{fig:AgenticApproach}. This paradigm offers several key benefits:

\noindent \textbf{Decomposition:} It breaks down the monumental task of documenting an entire system into smaller, logical steps (e.g., summarize code, retrieve existing docs, generate new ADRs, validate output).

\noindent \textbf{Specialization:} It assigns each step to a specialized agent (e.g., a Repo Summarizer, an ADR Generator) that is optimized for that specific task.

\noindent \textbf{Collaboration:} Agents pass information and artifacts to one another, ensuring that each stage builds upon a validated foundation. For instance, the summary from the Architecture Extractor provides crucial context for the Architecture Generator.

\noindent \textbf{Validation and Refinement:} It incorporates validator agents that act as quality control checkpoints. This allows for an iterative refinement process, significantly improving the accuracy and reliability of the final output.

\section{AKM Agents}\label{sec:agents}

\AgenticAKM~is a multi agent approach coordinated by a central Orchestrator and supervised by a human Architect as shown in Figure \ref{fig:AgenticApproach}. The Architect configures the initial workflow, monitors its execution, and can intervene to modify the flow if necessary. The Orchestrator manages the interaction between four specialized agent groups: Architecture Extractor, Retriever, Generator and Validator. In the rest of the section, we describe some of the Agents. It must be noted, this paper does not provide an exhaustive list of all potential agents within \AgenticAKM. The agents presented constitute an initial configuration that can be expanded and refined in subsequent research.

\subsection{Architecture Extractor Agents}
This group is responsible for parsing the code repository or other documentation, and creating high-level summaries that serve as the foundational architectural information for all other agents. Some of the agents can be:

\noindent \textbf{Repo Summarizer:} This agent analyzes the codebase to understand its overall architecture, primary components, and key functionalities. It uses an LLM to synthesize this analysis into a concise summary.

\noindent \textbf{Issue Summarizer:} This agent focuses on the project's issue tracker (e.g. Jira issues). It analyzes bug reports, feature requests, and developer discussions to extract architectural context and pain points, which can inform the AK generation process.

\noindent \textbf{Document Summarizer:} This agent ingests existing documentation within the repository (e.g. requirement docs) and uses an LLM to produce condensed summaries.

\subsection{Architecture Retriever Agents}
This group is responsible for locating and retrieving information from the project’s internal knowledge base or external sources. Its retriever agents use various tools—such as web search, vector retrieval, and knowledge base queries. Some of the agents can be:

\noindent \textbf{ADR Retriever:} This agent specializes in searching for existing ADRs. It utilizes a Vector Retrieval tool to perform semantic searches over a vector database of past decisions, helping in making and documenting future Design Decisions.

\noindent \textbf{Architecture Diagram Retriever:} This agent searches or retrieves existing architectural diagrams within the project's documentation or file system. It may also use image vector retrieval.

\noindent \textbf{Requirement Docs Retriever:} This agent uses file System Search and Vector Retrieval to locate and pull information from requirements documents, ensuring that generated architectural decisions align with specified project goals.

\subsection{Architecture Generator Agents}
This group is responsible for creating new architectural artifacts based on the context provided by the Extractor and Retriever agents. Some of the agents can be:

\noindent \textbf{ADR Generator:} This uses an LLM to draft new ADRs based on the repository summary and retrieved information. Each ADR is structured with standard sections like Title, Context, Decision, and Consequences.

\noindent \textbf{Diagram Generator:} This agent creates new architectural diagrams (e.g., UML, C4 models). It leverages a Image Generator Model to visualize the architecture described in the repository summary or a specific ADR.

\noindent \textbf{Specification Docs Generator:} This agent drafts technical specification documents for new components or services, using an LLM to ensure the documentation is detailed, clear, and consistent with the established architecture.

\subsection{Architecture Validator Agents}
This group acts as a quality assurance layer, scrutinizing the generated artifacts for accuracy, coherence, and compliance with organizational standards. Some of the agents can be:

\noindent \textbf{Repo Summary Validator:} This agent cross-references the summary created by the Extractor against the actual source code to ensure its accuracy and completeness. It uses an LLM to perform this comparative analysis.

\noindent \textbf{Document Summary Validator:} This agent checks the summaries of existing documents for factual correctness.

\noindent \textbf{ADR Validator:} This agent scrutinizes generated ADRs for logical consistency, format correctness, and overall quality. It may use Vector Retrieval to compare the new ADR against existing ones to check for redundancy or contradiction.

\noindent \textbf{Organization Policy Validator:} This agent ensures that generated architectural document comply with organizational best practices or predefined architectural principles stored in a knowledge base.

\section{Experiments}\label{sec:experiments}

\begin{table*}[th]
\centering
\begin{tabular}{llccccc}
\hline
\textbf{Source} & \textbf{Model} & \textbf{Relevance} & \textbf{Coherence} & \textbf{Completeness} & \textbf{Conciseness} & \textbf{Overall} \\
\hline
LLM        & GPT-5     & 3.8      & 3.8      & 3.7      & 3.5     & 3.3    \\
LLM        & Gemini    & 3.8      & 3.6      & 3.0      & 3.4     & 3.3    \\
Agent      & GPT-5     & 4.1      & \textbf{4.3}      & \textbf{3.9}      & 3.9     & 3.8    \\
Agent      & Gemini    & \textbf{4.3}      & 4.1      & 3.8      & \textbf{4.1}     & \textbf{3.9}    \\
\hline
\end{tabular}
\caption{User study results comparing Agentic vs. Baseline (LLM) approaches across two models. Scores are averaged over 29 repositories on a 5-point scale.}
\label{tab:user-study-results}
\end{table*}

To test the viability and effectiveness of \AgenticAKM, we made a instantiation of it with a system to automate the creation of ADRs from code repositories by dividing the task among specialized agents. Note that, ADR generation is an instantiation of AgenticAKM, not its full scope. The source code for all the experiments alongside the data used is available on GitHub \footnote{\url{https://github.com/sa4s-serc/AgenticAKM}}.

\subsection{Agentic ADR generation from repository}\label{subsec:adrFromRepo}
The workflow begins with a \textbf{Repo Summarizer Agent} analyzing the codebase to create a high-level summary. This summary is then validated by a \textbf{Summary Validator Agent}. If the summary is rejected, it is sent back to the Summarizer for refinement in a loop that runs up to three times.

Once the summary is approved, an \textbf{ADR Generator Agent} uses it to identify significant architectural decisions and draft corresponding ADRs. These drafts are scrutinized by an \textbf{ADR Validator     Agent} for correctness and quality. Similar to the summarization step, this triggers a refinement loop with the generator for up to three iterations if the ADRs are rejected. Finally the approved ADRs are saved.

The Orchestrator orchestrates the iterative process and calls the respective Agents when required. This agentic, iterative approach ensures that each stage builds upon a validated foundation, significantly improving the accuracy and relevance of the  generated ADRs. Note, for the user study, the workflow ran without human intervention.

\subsection{Evaluation Setup}
To evaluate the effectiveness of our proposed approach, we conducted a comparative user study. We designed the experiment to assess the quality of ADRs generated by two approaches across two different LLMs. We compared the two approaches for ADR generation:

\noindent \textbf{Baseline Approach:} This method involved extracting key files and components from a given repository and feeding this directly to an LLM in a single prompt to generate ADRs.

\noindent \textbf{Agentic Approach:} The agentic system detailed in subsection \ref{subsec:adrFromRepo}, uses a structured, iterative workflow involving summarizer, generator, and Validator agents to produce the final ADRs.

For the underlying LLMs, we selected 'Gemini-2.5-pro' and 'gpt-5', which were the top-ranked models on the LmArena leaderboard \cite{chiang2024chatbotarenaopenplatform} at the time of our experiment (October 5th, 2025). This resulted in four distinct experimental configurations:
\begin{itemize}[leftmargin=*]
    \item Baseline with Gemini
    \item Baseline with GPT
    \item Agentic with Gemini
    \item Agentic with GPT
\end{itemize}

\noindent \textbf{User Study Protocol:}\\
We performed the following steps in the user study:
\begin{itemize}[leftmargin=*]
    \item We initiated our user study by distributing a study form to recruit participants, targeting students and professionals with a background in software engineering and familiarity with ADRs.
    \item We received responses from 13 participants with 0 to 6 years of industry experience. They collectively submitted 29 unique code repositories in which they had direct expertise or had actively contributed. Python was the most prevalent language (17 repositories), followed by JavaScript (12 repositories), with repositories ranging from 1,000 to 350,000 lines.
    \item For each repository, we generated four distinct sets of ADRs, each corresponding to one of four experimental configurations. To ensure an unbiased evaluation, we employed a blind study design, anonymizing the Configurations ("config 1" to "config 4") so participants did not know which configuration produced which output.
    \item The participants were then asked to evaluate all four anonymized ADR sets for their repositories using a separate, structured feedback form.
    \item Following this, all responses were aggregated and analyzed to compare the efficacy of the different configurations.
\end{itemize}

\noindent The evaluation consisted of two parts:

\noindent \textbf{Quantitative Ratings:} Participants provided a star rating (from 1 to 5) for each set of ADRs based on four criteria: Relevance, Coherence, Completeness, Conciseness, and Overall Quality.

\noindent \textbf{Qualitative Feedback:} Participants also provided written comments detailing the strengths and weaknesses of the ADRs generated by each of the four configurations.

\subsection{Results}

The \textbf{quantitative} results in Table \ref{tab:user-study-results} show that the agentic approach consistently outperforms the LLM only approach across all evaluation metrics. Agentic approach attained the highest overall quality score of 3.9. In comparison, the LLM-only configurations scored 3.3 overall, with notably lower Completeness ratings, indicating occasional omissions and limited reasoning depth. The agentic approach also maintained strong Conciseness (3.9–4.1) and Relevance (4.3-4.1) scores, reflecting an effective balance between brevity and informational richness.

The \textbf{qualitative} feedback from participants further reinforces these findings. While some users acknowledged the LLM-only outputs as "to the point" or praised them for "good coverage of the whole repo," others criticized them as "very wordy" or lacking structure. In contrast, the outputs generated via the agentic approach were praised more with comments as "very structured and clear" and more reflective of "actual architectural reasoning." One participant noted that the agentic ADRs “actually captured different underlying decisions,” whereas the LLM-only outputs appeared "very abstract and generic."
We also observed that while language of repository didn’t play a major role in the quality of ADRs, languages like Java had higher quality ADRs generated with LLMs over Agentic approach

The results show that \AgenticAKM~significantly improves ADR quality over simple LLM calls, producing more complete, concise, and contextually accurate documentation. This highlights the potential of Agentic AI for enhancing AKM.

\section{Related Works}\label{sec:RelatedWorks}

Recent advances in GenAI have begun reshaping software architecture research. Esposito et al. \cite{esposito2025generativeaisoftwarearchitecture} mapped the emerging use of LLMs in architectural reconstruction, documentation, and decision support, while Ivers et al. \cite{GenAiAutomationArch2025} examined which architectural activities are realistically automatable, stressing the enduring need for human governance. Empirical studies such as Dhar et al. \cite{dhar2024llmsgeneratearchitecturaldesign} explored LLMs’ ability to generate architectural decisions from context, revealing model and prompt dependent variability. Similarly, Manjula and Dube \cite{GenAiArchDiagram} demonstrated the use of LLMs for creating and interpreting architecture diagrams. These efforts show that LLMs can support AKM but lack its integration into repository.

On the other hand, Agentic AI in being heavily used to tackle software engineering problems, with some research focusing on foundational design patterns for building them \cite{liu2025agent}. For example, Wadhwa et al. \cite{wadhwa2024masai} and Bouzenia et al. \cite{bouzenia2025understandingsoftwareengineeringagents} analyzed distributed AI agents that decompose software development work into collaborative reasoning cycles, exposing both potential and challenges.

Agentic systems have also been applied in software architecture, an intersection broadly explored by Vaidhyanathan et al. \cite{vaidhyanathan2025software}. Díaz-Pace et al. \cite{ArchExplorationLlmAgent} proposed ReArch, a reflective LLM-based approach in which autonomous agents explore architectural design alternatives and reason about trade-offs. Similarly, Li et al. \cite{li2025maadautomatesoftwarearchitecture} introduced MAAD (Multi-Agent Automated Architecture Design), where specialized agents collaborate to synthesize and evaluate new architectures.

While prior research has primarily focused on design synthesis, emphasizing the creation of new architectural solutions, our work instead applies agentic reasoning to AKM automating the extraction, refinement, and documentation of architecture knowledge from existing software systems, and advancing knowledge recovery and preservation rather than design generation.

\section{Conclusion and Future Works}\label{sec:conclusion}

This paper presented a novel agentic approach for AKM, \AgenticAKM. By decomposing architecture recovery into specialized Extractor, Retriever, Generator, and Validator agents, the approach overcomes the limitations of monolithic LLM calls. A user study reveals that the approach produces better ADRs form code repositories. The findings validate that \AgenticAKM~offers a robust and effective methodology for automating AKM.

Future work will expand the approach to generate additional artifacts, such as C4 diagrams, and enhance human-agent collaboration through an architect in the loop model. Longitudinal industrial studies will assess scalability and long term impact, while new agents will be developed to process unstructured, multi modal data, capturing richer contextual design knowledge.

\begin{acks}
Students and practitioners who contributed to the study.
\end{acks}

\bibliographystyle{unsrt}
\bibliography{reference}

@String{Springer = "Springer-Verlag" }

@misc{wu2023autogenenablingnextgenllm,
      title={AutoGen: Enabling Next-Gen LLM Applications via Multi-Agent Conversation}, 
      author={Qingyun Wu and Gagan Bansal and Jieyu Zhang and Yiran Wu and Beibin Li and Erkang Zhu and Li Jiang and Xiaoyun Zhang and Shaokun Zhang and Jiale Liu and Ahmed Hassan Awadallah and Ryen W White and Doug Burger and Chi Wang},
      year={2023},
      eprint={2308.08155},
      archivePrefix={arXiv},
      primaryClass={cs.AI},
      url={https://arxiv.org/abs/2308.08155}, 
}

@misc{esposito2025generativeaisoftwarearchitecture,
      title={Generative AI for Software Architecture. Applications, Challenges, and Future Directions}, 
      author={Matteo Esposito and Xiaozhou Li and Sergio Moreschini and Noman Ahmad and Tomas Cerny and Karthik Vaidhyanathan and Valentina Lenarduzzi and Davide Taibi},
      year={2025},
      eprint={2503.13310},
      archivePrefix={arXiv},
      primaryClass={cs.SE},
      url={https://arxiv.org/abs/2503.13310}, 
}

@INPROCEEDINGS{GenAiAutomationArch2025,
  author={Ivers, James and Ozkaya, Ipek},
  booktitle={2025 IEEE 22nd International Conference on Software Architecture Companion (ICSA-C)}, 
  title={Will Generative AI Fill the Automation Gap in Software Architecting?}, 
  year={2025},
  volume={},
  number={},
  pages={41-45},
  doi={10.1109/ICSA-C65153.2025.00014}}

@misc{dhar2024llmsgeneratearchitecturaldesign,
      title={Can LLMs Generate Architectural Design Decisions? -An Exploratory Empirical study}, 
      author={Rudra Dhar and Karthik Vaidhyanathan and Vasudeva Varma},
      year={2024},
      eprint={2403.01709},
      archivePrefix={arXiv},
      primaryClass={cs.SE},
      url={https://arxiv.org/abs/2403.01709}, 
}

@article{GenAiArchDiagram,
author = {Manjula, Nishchai and Dube, Akhilesh},
year = {2024},
month = {12},
pages = {3330-3336},
title = {Harnessing generative AI to create and understand architecture diagrams},
volume = {13},
journal = {International Journal of Science and Research Archive},
doi = {10.30574/ijsra.2024.13.2.2601}
}

@INPROCEEDINGS{ArchExplorationLlmAgent,
  author={Diaz-Pace, J. Andrés and Tommasel, Antonela and Capilla, Rafael and Ramírez, Yamid E.},
  booktitle={2025 IEEE 22nd International Conference on Software Architecture Companion (ICSA-C)}, 
  title={Architecture Exploration and Reflection Meet LLM-based Agents}, 
  year={2025},
  volume={},
  number={},
  pages={1-5},
  doi={10.1109/ICSA-C65153.2025.00015}
}

@misc{li2025maadautomatesoftwarearchitecture,
      title={MAAD: Automate Software Architecture Design through Knowledge-Driven Multi-Agent Collaboration}, 
      author={Ruiyin Li and Yiran Zhang and Xiyu Zhou and Peng Liang and Weisong Sun and Jifeng Xuan and Zhi Jin and Yang Liu},
      year={2025},
      eprint={2507.21382},
      archivePrefix={arXiv},
      primaryClass={cs.SE},
      url={https://arxiv.org/abs/2507.21382}, 
}

@inproceedings{wadhwa2024masai,
title={{MASAI}: Modular Architecture for Software-engineering {AI} Agents},
author={Nalin Wadhwa and Atharv Sonwane and Daman Arora and Abhav Mehrotra and Saiteja Utpala and Ramakrishna B Bairi and Aditya Kanade and Nagarajan Natarajan},
booktitle={NeurIPS 2024 Workshop on Open-World Agents},
year={2024},
url={https://openreview.net/forum?id=NSINt8lLYB}
}

@misc{bouzenia2025understandingsoftwareengineeringagents,
      title={Understanding Software Engineering Agents: A Study of Thought-Action-Result Trajectories}, 
      author={Islem Bouzenia and Michael Pradel},
      year={2025},
      eprint={2506.18824},
      archivePrefix={arXiv},
      primaryClass={cs.SE},
      url={https://arxiv.org/abs/2506.18824}, 
}

@misc{chiang2024chatbotarenaopenplatform,
      title={Chatbot Arena: An Open Platform for Evaluating LLMs by Human Preference}, 
      author={Wei-Lin Chiang and Lianmin Zheng and Ying Sheng and Anastasios Nikolas Angelopoulos and Tianle Li and Dacheng Li and Hao Zhang and Banghua Zhu and Michael Jordan and Joseph E. Gonzalez and Ion Stoica},
      year={2024},
      eprint={2403.04132},
      archivePrefix={arXiv},
      primaryClass={cs.AI},
      url={https://arxiv.org/abs/2403.04132}, 
}

@article{sapkota2025ai,
  title={Ai agents vs. agentic ai: A conceptual taxonomy, applications and challenges},
  author={Sapkota, Ranjan and Roumeliotis, Konstantinos I and Karkee, Manoj},
  journal={arXiv preprint arXiv:2505.10468},
  year={2025}
}

@inproceedings{vaidhyanathan2025software,
  title={Software Architecture in the Age of Agentic AI},
  author={Vaidhyanathan, Karthik and Muccini, Henry},
  booktitle={European Conference on Software Architecture},
  pages={41--49},
  year={2025},
  organization={Springer}
}

@article{liu2025agent,
  title={Agent design pattern catalogue: A collection of architectural patterns for foundation model based agents},
  author={Liu, Yue and Lo, Sin Kit and Lu, Qinghua and Zhu, Liming and Zhao, Dehai and Xu, Xiwei and Harrer, Stefan and Whittle, Jon},
  journal={Journal of Systems and Software},
  volume={220},
  pages={112278},
  year={2025},
  publisher={Elsevier}
}

@article{xi2025rise,
  title={The rise and potential of large language model based agents: A survey},
  author={Xi, Zhiheng and Chen, Wenxiang and Guo, Xin and He, Wei and Ding, Yiwen and Hong, Boyang and Zhang, Ming and Wang, Junzhe and Jin, Senjie and Zhou, Enyu and others},
  journal={Science China Information Sciences},
  volume={68},
  number={2},
  pages={121101},
  year={2025},
  publisher={Springer}
}

@article{he2025llm,
  title={LLM-Based Multi-Agent Systems for Software Engineering: Literature Review, Vision, and the Road Ahead},
  author={He, Junda and Treude, Christoph and Lo, David},
  journal={ACM Transactions on Software Engineering and Methodology},
  volume={34},
  number={5},
  pages={1--30},
  year={2025},
  publisher={ACM New York, NY}
}

@article{wang2025agents,
  title={Agents in software engineering: Survey, landscape, and vision},
  author={Wang, Yanlin and Zhong, Wanjun and Huang, Yanxian and Shi, Ensheng and Yang, Min and Chen, Jiachi and Li, Hui and Ma, Yuchi and Wang, Qianxiang and Zheng, Zibin},
  journal={Automated Software Engineering},
  volume={32},
  number={2},
  pages={1--36},
  year={2025},
  publisher={Springer}
}

\end{document}